\documentstyle[epsf,preprint,aps,prb,floats,tighten]{revtex}
\begin{document}
\draft
\preprint{January 30, 1997}


\title{First-principles theory of structural phase transitions for
perovskites: competing instabilities}

\author{David Vanderbilt and W. Zhong}

\address{Department of Physics and Astronomy
  Rutgers University, Piscataway, NJ 08855-0849}

\date{January 30, 1997}
\maketitle


\begin{abstract}
We extend our previous first-principles theory for perovskite
ferroelectric phase transitions to treat also antiferrodistortive
phase transitions.  Our approach involves construction of a model
Hamiltonian from a Taylor expansion, first-principles calculations
to determine expansion parameters, and Monte Carlo simulations to
study the resulting system.  We apply this approach to three cubic
perovskite compounds, SrTiO$_3$, CaTiO$_3$, and NaNbO$_3$, that are
known to undergo antiferrodistortive phase transitions.  We
calculate their transition sequences and transition temperatures at
the experimental lattice constants.  For SrTiO$_3$, we find our
results agree well with experiment.  For more complicated compounds
like CaTiO$_3$ and NaNbO$_3$, which can have many different
structures with very similar energy, the agreement is somewhat less
satisfactory.
\end{abstract}


\bigskip
\pacs{Keywords: structural phase transitions, ferroelectrics, 
SrTiO$_3$, CaTiO$_3$, NaNbO$_3$}

\narrowtext

\section{Introduction}

Perovskite materials are of considerable interest both for
fundamental reasons and for their many actual and potential
technological applications.  The great fascination of the cubic
perovskite structure is that it can readily display a variety of
structural phase transitions, ranging from non-polar
antiferrodistortive (AFD) to ferroelectric (FE) and
antiferroelectric (AFE) in nature.\cite{lines}  The competition
between these different instabilities evidently plays itself out in
a variety of ways, depending on the chemical species involved,
leading to the unusual variety and richness of the observed
structural phase diagrams.  For example, as temperature is reduced,
BaTiO$_3$ undergoes a series of FE phase transitions, while
SrTiO$_3$ has a single AFD transition.  More extreme examples are
NaNbO$_3$ and BaZrO$_3$; the former has a series of six
transitions, while the latter stays cubic down to zero
temperature.  Another appealing property of these cubic perovskites
is that all of the structural phase transitions involve only small
distortions from the ideal cubic structure, the typical distortion
being less than 5\% of the lattice constant.  This simplifies the
theoretical treatment considerably.  The ample experimental data on
these compounds also provide many insights and opportunities for
checking the accuracy of theoretical calculations.

It is no wonder that there have been many theoretical attempts to
study these compounds. Previous phenomenological model Hamiltonian
approaches\cite{dove,pytt,cowl,pytt2} have largely been limited by
oversimplification and ambiguities in interpretation of experiment,
while empirical \cite{bilz} and non-empirical pair-potential methods
\cite{boyer} have not offered high enough accuracy.  Recently,
advances in density-functional techniques have made possible
first-principles investigations of such perovskite compounds.  Such
calculations have proven capable of providing accurate structural
properties and FE distortions for perovskites at zero
temperature.\cite{cohen,singh,king1}

Recently, a thermodynamic theory based upon such first-principles
calculations was developed to study the finite-temperature
properties of BaTiO$_3,$\cite{zhong2,zhong2b} and predicted the
correct transition sequence and fairly accurate transition
temperatures.  This thermodynamic approach involves three steps:
(i) constructing an effective Hamiltonian to describe the important
degrees of freedom of the system; (ii) determining all the
parameters of this effective Hamiltonian from high-accuracy {\it
ab-initio} LDA calculations; and (iii) performing Monte Carlo
simulations to determine the phase transformation behavior of the
resulting system.  A similar approach was also successfully applied
to PbTiO$_3$ by Rabe and Waghmare.\cite{rabeuw}

The construction of the effective Hamiltonian is carried out in
view of the special structure properties of cubic perovskite
compounds.  At higher temperature, the cubic perovskite compounds
ABO$_3$ have a simple cubic structure with O atoms at the face
centers and metal atoms A and B at the cube corner and body center,
respectively.  The two most common instabilities result from the
softening of either a polar zone-center phonon mode, leading to a
FE phase,
or the softening of a non-polar zone-boundary mode involving
rotations of oxygen octahedra, leading to an AFD phase.  (In some
cases a zone-boundary polar mode may also occur, leading to an AFE
phase.)  In our previous thermodynamic theory for BaTiO$_3$, we
assumed FE and strain distortions would be the only important
degrees of freedom of the system. In other words, all other
distortions are assumed to be much higher in energy.  This is true
for BaTiO$_3$, but not true for cubic perovskites in general.  As
shown in our recent first-principle calculations,\cite{zhong3} most
cubic perovskite compounds may also undergo AFD transitions. To
study these compounds, we need to extend our theory to include AFD
distortions among the low-energy distortions. This extended theory
would also allow us to study the interaction between FE and AFD
instabilities. However, because the AFD distortion is a
zone-boundary distortion without a clear corresponding zone-center
mode, the extension is not trivial.

The rest of the paper is organized as follows.  In Sec.~II, we go
through the detailed procedure for the construction of the
effective Hamiltonian with the AFD distortion included.  In
Sec.~III, we describe our first-principles calculations and the
determination of the expansion parameters for the three compounds
SrTiO$_3$, CaTiO$_3$, and NaNbO$_3$.  In Sec.~IV, we report our
calculated transition temperatures, phase sequences, and order
parameters for those three compounds.  We also identify the
differences between the correlation functions of the FE and AFD
local modes in SrTiO$_3$.  Sec.~V concludes the paper.

\section{Construction of the Hamiltonian}

\subsection{Local modes for AFD distortion}

In our previous development,\cite{zhong2b} we argued that the total
energy of a cubic perovskite can be well approximated by a
low-order Taylor expansion over all the relevant low-energy
distortions, specifically FE distortions and strain.  The FE
distortions are represented by local modes, whose arrangement will
reproduce the FE soft phonon modes throughout the Brillouin zone
(BZ).  To extend the theory to include the AFD distortions, we need
to construct a new set of local modes to represent the lowest AFD
modes over the whole BZ, or at least over the portion of the BZ
where the energy change due to the AFD distortions is either
negative, or positive but small.  The AFD mode typically has the
lowest energy at the zone-boundary R (0.5, 0.5, 0.5)$2\pi/a_0$
and M [(0.5, 0.5, 0)$2\pi/a_0$, etc.] points, while near the zone
center $\Gamma$ the energy is very high.  So it is necessary to
choose local modes that will accurately reproduce the potentially
soft modes in the vicinity of the R and M points.

The rotation of an isolated oxygen octahedron can be represented
by a pseudovector passing through its center.  Assuming
the origin of coordinates at the center of the octahedron, a
pseudovector with polarization $\hat{\bf z}$ involves displacements
$\pm 0.5a_0 \hat{\bf y} $ for oxygen atoms at ($\pm a_0$/2,0,0) and
displacements $\mp 0.5a_0 \hat{\bf x}$ for atoms at (0,$\pm
a_0$/2,0).  Here $a_0$ is the lattice constant of the ideal cubic
perovskite.  In the case of the ABO$_3$ perovskite crystal, we can
represent octahedral rotation using pseudovectors sitting on the
center of each octahedron, i.e., on the B atoms.  However,
neighboring octahedra share oxygen atoms, so that some continuity
conditions would have to be imposed if we were to insist that the
displacement of a given oxygen be consistently described by both
neighboring pseudovectors.  With such constraints the neighboring
pseudovectors would no longer be independent of one another,
leading to potential problems in the implementation of the Monte
Carlo simulations.

To avoid such problems, we simply construct a set of ``virtual''
pseudovectors ${\bf a}_i$ which are independent of each other,
and let the actual oxygen displacements be the superposition of the
displacements that would result from these.  To be precise, let
${\bf a}_i \equiv {\bf a}({\bf R}_i)$ denote the pseudovector
centered on the B atom of unit cell $i$ (position vector ${\bf
R}_i$), so that each oxygen atom is shared by two pseudovectors.
The physical displacement of the oxygen atom shared by ${\bf a}_i$
and ${\bf a}_j$ is then given by
\begin{equation}
\Delta {\bf r} = \frac {a_0}{2} \hat{\bf R}_{ij} \times ({\bf a}_i-
{\bf a}_j) \; ,
\label{ro}
\end{equation}
where ${\bf R}_{ij} = {\bf R}_i - {\bf R}_j$,
$\hat{\bf R}_{ij} = {\bf R}_{ij} /|{\bf R}_{ij}|$.
The AFD soft modes of interest at R and M are then easily
represented by the corresponding pattern of pseudovectors.  For
example, choosing ${\bf a}(l{\bf x}+m{\bf y}+n{\bf z})=(-)^{l+m}
\hat{\bf z}$ reproduces one of the M-point modes polarized along
$\hat{\bf z}$.  (Here, ${\bf x}=a_0\hat{\bf x}$, etc.)
Other possible choices of the pattern of
pseudovectors correspond to other modes which are probably higher
in energy, but possibly still relevant for some materials.  For
example, choosing ${\bf a}(l{\bf x}+m{\bf y}+n{\bf z})=(-)^l
\hat{\bf y}$ corresponds to an X-point mode that can be regarded
as either of AFD character polarized along $\hat{\bf y}$, or of AFE
character polarized along $\hat{\bf z}$.  Finally, note that
choosing ${\bf a}({\bf R}_i)$ constant ($\Gamma$-point arrangement)
gives rise to no displacements whatever.

In view of this last point, it is important realize that the ${\bf
a}_i$ themselves do not have direct physical meaning; only
differences between adjacent ${\bf a}_i$'s are physical.  Adding a
constant to all ${\bf a}_i$'s will not change the physical
configuration of the system. So any physical distortions can be
mapped to infinitely many pseudovector arrangements, but any
pseudovector arrangement only corresponds to one specific physical
distortion.  Because only the pseudovector differences between
sites have physical meaning, the Hamiltonian should be expanded in
terms of these differences, not the pseudovectors themselves.
Using this approach, we reduce the number of degrees of freedom
associated with oxygen displacements perpendicular to the O--B
bonds from six to three, and raise the symmetry of the system
considerably.  As a result, the Taylor expansion is significantly
simplified.

The two other low-energy distortions, the FE and elastic
distortions, are treated as in Ref.\onlinecite{zhong2b}.  Briefly,
for each unit cell of the ABO$_3$ perovskite structure, we define a
FE local-mode centered on the B site, and a displacement local mode
centered on the A site.  The former is chosen in such a way that a
uniform superposition of FE local modes reproduces the soft TO mode
obtained by diagonalizing the force-constant matrix at the
Brillouin zone center.  The quantities ${\bf f}_i \equiv {\bf f}
({\bf R}_i) $ and ${\bf u}_i$ are the vector amplitudes of the FE
and translational local modes, respectively, in the $i$th unit
cell.  (Note the difference in notation between
Ref.\onlinecite{zhong2b} and this paper.\cite{notation})
Inhomogeneous strains $\eta$ are expressed in terms of differences
between ${\bf u}_i$ in neighboring cells, and we add six extra
degrees of freedom to describe the homogeneous strain.  Thus, the
total energy $E^{\rm tot}$ depends on the set of variables \{${\bf
f}_i, {\bf a}_i, {\bf u}_i$\} and the homogeneous strain, and is
expanded in a Taylor series in terms of these quantities.  The
expansion terms can be divided into four kinds, those involving
the FE local modes ${\bf f}_i$ alone, the AFD modes ${\bf
a}_i$ alone, the strain variable ${\bf u}_i$ alone, and the coupling
between them,
\begin{equation}
E^{\rm tot} = E^{\rm F}(\{ {\bf f} \} ) + E^{\rm A} (\{ {\bf a} \} )
+ E^{\rm E} (\{ {\bf u} \} )+ E^{\rm int} (\{ {\bf f,a,u} \} )\; .
\end{equation}
The part of the energy involving the FE local modes alone, $E^{\rm
F}(\{{\bf f}\})$, contains the on-site self energy, dipole-dipole
interactions, and short-range residual interactions.  Their forms
have been given by equations in Ref.\onlinecite{zhong2b}: Eq.\ (3)
in Sec.~II.B, Eq.\ (7) in Sec.~II.C, and Eq.\ (9) in Sec.~II.D,
respectively.\cite{notation} The energy due to ${\bf u}_i$ alone,
$E^{\rm E}(\{{\bf u}\})$, is just the elastic energy, and its form
has been given by Eq.\ (11) of Ref.\onlinecite{zhong2b}.  Also, the
energy terms representing coupling between the ${\bf f}_i$ and
${\bf u}_i$ have been given in Eq.\ (14) of Sec.~II.F of
Ref.\onlinecite{zhong2b}.

It remains to present here the energy terms involving solely the AFD
local modes ${\bf a}_i$ and those representing the coupling of the
${\bf a}_i$ to the ${\bf f}_i$ and ${\bf u}_i$. Their expressions are
presented in the following.

\subsection{AFD energy terms}

The AFD local modes ${\bf a}_i$ are nonpolar and involve no dipole
moment, so long-range dipole-dipole interactions need not be
considered, unlike for FE local modes.  Recalling that only the
differences of the ${\bf a}_i$ between neighboring sites are
physical, it is not appropriate to separate energy contributions
into on-site and inter-site interactions, as we did for the ${\bf
f}_i$.\cite{zhong2b} Instead, we separate the interaction into
harmonic and higher-order contributions,
\begin{equation}
E^{\rm A} (\{ {\bf a} \} ) = E^{\rm A, harm} (\{ {\bf a} \} ) +
E^{\rm A, anharm} (\{ {\bf a} \} ) \;.
\end{equation}
In principle, all the AFD energy terms should be expanded in terms
of the $\Delta{\bf r}$ expressed through Eq.\ (\ref{ro}).  For
intersite interactions, this would become very complicated because
of the low crystal symmetry at the O sites.  However, for harmonic
terms, the expression can be simplified by expansion in terms of
the ${\bf a}_i$ directly with certain conditions enforced.  In this
case, we can write
\begin{equation}
E^{\rm A, harm} (\{ {\bf a} \} ) =
  \frac{1}{2} \sum_{ij} \sum_{\alpha\beta}
J^{\rm A}_{ij,\alpha\beta} a_{i\alpha} a_{j\beta} \; .
\label{eqshort}
\end{equation}
Here, $\alpha$ and $\beta$ denote Cartesian components, and $J^{\rm
A}_{ij,\alpha\beta}$ is a function of ${\bf R}_{ij}$ and should
decay very fast with increasing $|{\bf R}_{ij}|$.  We need to
impose conditions on the $J^{\rm A}_{ij,\alpha\beta}$ reflecting
the fact that the dependence of the energy on the ${\bf a}_i$ is
only through differences between neighboring sites [Eq.\ (\ref{ro})].
The appropriate conditions are
\begin{equation}
\sum_{j\,\in\,{\rm plane}\;m} J^{\rm A}_{ij,\alpha\beta}  = 0 \; ,
\label{cond1}
\end{equation}
where the sum is over sites $j$ such that $R_{ij,\beta}=ma_0$.
This reflects the fact that if we make a change ${\bf a}_j
\rightarrow {\bf a}_j + c\hat{\bf z}$ for all the pseudovectors
in an $x$-$y$ plane at a distance of $m$ unit cells away from the
site $i$, the resulting the ``force'' on the pseudovector on site
$i$ should vanish.  It can be shown that with these conditions
enforced, the interaction energy is only related to pseudovector
difference between adjacent sites.

The description in terms of the ${\bf a}_i$ directly makes it
possible to simplify the interaction matrix $J^{\rm
A}_{ij,\alpha\beta}$ by symmetry, since ${\bf a}_i$ is centered on
the high-symmetry sites.  For a cubic lattice, we have
\begin{eqnarray}
{\hbox{\rm on-site}: \quad} & J^{\rm A}_{ii,\alpha\beta} & =
       2 \kappa_2^{\rm A}\delta_{\alpha\beta} \; ,
                                            \nonumber\\
{\rm 1st \; nn: \quad} & J^{\rm A}_{ij,\alpha\beta} & = [j^{\rm A}_1 +
  (j^{\rm A}_2-j^{\rm A}_1)\theta_{ij,\alpha}]\delta_{\alpha\beta} \; ,
                                            \nonumber\\
{\rm 2nd \; nn: \quad} & J^{\rm A}_{ij,\alpha\beta}  & =
       [j^{\rm A}_4 + (j^{\rm A}_3-j^{\rm A}_4) \theta_{ij,\alpha}]
       \delta_{\alpha\beta}
                                            \nonumber\\
   & & \qquad + j^{\rm A}_5 \theta_{ij,\alpha} \theta_{ij,\beta}
            (1-\delta_{\alpha\beta}) \; ,
                                            \nonumber\\
{\rm 3rd \; nn: \quad} & J^{\rm A}_{ij,\alpha\beta} & =
 j^{\rm A}_6 \delta_{\alpha\beta} + j^{\rm A}_7 \theta_{ij,\alpha}
 \theta_{ij,\beta} (1-\delta_{\alpha\beta}) \; ,
                                            \nonumber\\
{\rm 4th \; nn: \quad} & J^{\rm A}_{ij,\alpha\beta} & =
 j^{\rm A}_8 \theta_{ij,\alpha} \delta_{\alpha\beta} \; ,
\label{bigeq}
\end{eqnarray}
where $\theta_{ij,\alpha}$=1 if $R_{ij}$ has a non-zero $\alpha$
component and 0 otherwise.  We include in-plane interactions
($R_{i,\alpha}= R_{j,\alpha}$) to 4th neighbor, since this kind of
interaction is much stronger than other interactions.  The
conditions Eq.\ (\ref{cond1}) can then be simplified to
\begin{eqnarray}
\kappa^{\rm A}_2 + 2 j^{\rm A}_1 + 2 j^{\rm A}_4
                       + 2 j^{\rm A}_8 & = & 0 \; , \nonumber\\
j^{\rm A}_2 + 4 j^{\rm A}_3 + 4 j^{\rm A}_6  & = & 0 \; .
\end{eqnarray}
Thus, the complicated harmonic intersite interaction matrix 
for AFD local distortions can be determined
from seven independent interaction parameters.

The structural phase transition problem is intrinsically an
anharmonic problem.  Since the harmonic modes may be unstable, it
is necessary to introduce higher order terms.  For simplicity, we
first only consider on-site anharmonic contributions associated
with oxygen atoms. Because of the tetragonal symmetry on the O
sites, the lowest anharmonic terms are of fourth order.  Since each
oxygen involves two nearest neighbor AFD pseudovectors, this
quartic term will take the form
\begin{eqnarray}
E^{\rm A, quart} (\{ {\bf a} \} ) & = & \sum_i
\sum_{{\bf d}=\pm {\bf x}} \alpha^{\rm A} \left\{ [ a_y ({\bf R}_i )
-a_y({\bf R}_i + {\bf d}) ]^4 + [ a_z ({\bf R}_i )-a_z({\bf R}_i+
{\bf d})]^4
\right\}  \nonumber\\
 & + & \sum_i \sum_{{\bf d}=\pm {\bf x}} \gamma^{\rm A}
 [ a_y ({\bf R}_i )-a_y({\bf R}_i + {\bf x}) ]^2 
 [ a_z ({\bf R}_i )-a_z({\bf R}_i + {\bf x}) ]^2 
 \nonumber\\
& + & {\rm \;cyclic\;permutations\;}.
\end{eqnarray}
Here, ${\bf x} = a_0 \hat{\bf x}$, and $\alpha^{\rm A}$ and
$\gamma^{\rm A}$ are parameters to be determined from
first-principles calculations.

In our previous work on BaTiO$_3$ the intersite FE interactions
have been expanded only up to harmonic order.  For AFD interactions
the corresponding approximation would be to truncate the
interactions between the AFD-induced displacements of different
oxygen atoms to harmonic order.  [Such terms are already included
in Eq.\ (\ref{bigeq}).] We find this approximation to be
satisfactory for those compounds with weak distortions, as in the
case of BaTiO$_3$ or SrTiO$_3$.  For CaTiO$_3$, the AFD distortion
is very large and the transition temperature is around 2000K.  In
this case, we find it necessary to include more complicated
anharmonic terms, such as third-order intersite interaction terms,
for the AFD distortions. In fact, such terms turn out to be
responsible for inducing a displacement component corresponding to
an X-point phonon (with both O and Ca character) in CaTiO$_3$.
For NaNbO$_3$, although the distortion is not as strong (the
highest transition temperature is around 700K), there are many
structures with very close free energy.  We find that inclusion of
the third-order AFD terms does have a noticeable effect for these
compounds, so we include these third order interaction terms for
CaTiO$_3$ and NaNbO$_3$.

We consider only those third-order interactions between AFD modes on
two or three neighboring lattice sites.  We can follow the treatment of
the harmonic intersite interactions by listing all the possible
interactions and using symmetry arguments to eliminate forbidden
terms.  Following this approach leads to three kind of terms, and
we would need three more parameters to fully specify the Hamiltonian.
Since the third-order terms are relatively weak and the exact
determination of the three parameters is costly, we investigate the
relations between these three parameters, and use a simple argument
to combine the three terms to form a single new term with only one
free parameter to determine.

The AFD interactions involve only the displacement of oxygen
atoms.  The strongest energy difference is associated with the
distortion of oxygen octahedra, or the change of the length of a
nearest-neighbor O--O bond $\Delta l$. We can start by analyzing
$\Delta l$ for two oxygens at (0,0,$a$/2) and ($a$/2,0,0). We
approximate the total-energy change as solely due to $\Delta l$.
Expanding it as a function of the rotation vectors ${\bf a}(i)$ up
to the third order, we obtain the desired third-order intersite
terms.  Using the short-hand notations
\begin{eqnarray}
 {\bf a'}(x) & \equiv & {\bf a}({\bf R}_i + {\bf x})
-{\bf a}({\bf R}_i) \;,
\nonumber \\
 {\bf a'}(-x) & \equiv & {\bf a}({\bf R}_i - {\bf x})
-{\bf a}({\bf R}_i) \;,
\end{eqnarray}
the cubic coupling term involving one O--O bond is
\begin{equation}
B_3 [a'_y (z) - a'_y(x)][a'_x(z)+a'_z(x)]^2 \; .
\end{equation}
The other 11 nearest-neighbor O--O bonds will give rise to 11 other
terms and the overall total-energy contribution can be expressed as
$E^{\rm A, cub}  =  B_3 \sum_i W_i$, where
\begin{eqnarray}
W_i = \{ \, & & + [ a'_x(y) -a'_x (z)][a'_y(z)+a'_z(y)]^2 \nonumber \\
 & & - [ a'_x (-y) - a'_x(z)][a'_y(z)-a'_z(-y)]^2  \nonumber \\
 & & - [ a'_x (y) - a'_x(-z)][a'_y(-z)-a'_z(y)]^2  \nonumber \\
 & & + [ a'_x(-y) -a'_x (-z)][a'_y(-z)+a'_z(-y)]^2  \nonumber \\
 & & \qquad + \; \hbox{\rm cyclic permutations} \,\}  \; .
\end{eqnarray}
This assumption that $\Delta l$ is solely responsible for the cubic
intersite interactions significantly simplifies the energy expression
and reduces the number of expansion parameters from three to one.

\subsection{Coupling energy}

There are three kind of coupling energy terms: those between FE and
elasticity, between FE and AFD, and between AFD and elasticity,
\begin{equation}
 E^{\rm int} = E^{\rm F-E} + E^{\rm F-A} + E^{\rm A-E}
\end{equation}
For simplicity, we consider only on-site couplings.  The coupling
between ${\bf f}_i$ and ${\bf u}_i$ ($E^{\rm F-E}$) has been given by
Eq.\ (14) in Sec.~II.F of Ref.\onlinecite{zhong2b}.\cite{notation}
Here, we expressions for $E^{\rm A-E}$ and $E^{\rm F-A}$.

The coupling between elasticity and AFD modes at lowest order can
be written as
\begin{eqnarray}
E^{\rm E-A}(\{ {\bf a} \},\{\eta \}) & =
& \frac{1}{2} \sum_{il\alpha\beta}
B_{l \alpha\beta x} \eta_l({\bf R}_i)
[\bar{a}_{i,\alpha}({\bf x}) \bar{a}_{i,\beta}({\bf x})+
\bar{a}_{i,\alpha}(-{\bf x}) \bar{a}_{i,\beta}(-{\bf x})]
\nonumber\\
& & \qquad + \; \hbox{\rm cyclic permutations} \;,
\end{eqnarray}
where $\bar{a}_{i,\alpha}({\bf d}) \equiv a_{\alpha} ({\bf
R}_i+{\bf d}) - a_{\alpha}({\bf R}_i)$ and $\eta_l({\bf R}_i)$ is
the six-component local strain tensor in Voigt notation ($\eta_1 =
e_{11}$, $\eta_4 = 2 e_{23}$).  $\eta_l({\bf R}_i)$ can be
expressed as a function of ${\bf u}_i$ following Sec.~II.F of
Ref.\onlinecite{zhong2b}.\cite{notation}   $B_{l\alpha\beta\gamma}$
is a high order coupling tensor.  Because of the symmetry, there
are only four independent coupling constants in
$B_{l\alpha\beta\gamma}$,
\begin{eqnarray}
B_{1yyx} = B_{1zzx} = B_{2xxy} = B_{2zzy} =B_{3xxz} =B_{3yyz}
\;,\nonumber\\ 
B_{2yyx} = B_{3zzx} = B_{1xxy} = B_{3zzy} =B_{1xxz} =B_{2yyz}
\;,\nonumber\\ 
B_{3yyx} = B_{2zzx} = B_{3xxy} = B_{1zzy} =B_{2xxz} =B_{1yyz}
\;,\nonumber\\ 
B_{4yzx} = B_{4zyx} = B_{5xzy} = B_{5zxy} =B_{6xyz} =B_{6yxz}
\; .
\end{eqnarray}
All other elements are zero.

The lowest-order coupling between ${\bf a}_i$ and ${\bf f}_i$ is
linear with both ${\bf a}_i$ and ${\bf f}_i$.  It takes the form
\begin{eqnarray}
 E_1^{\rm F-A} & = &
\sum_i G_{xy} f_{x}({\bf R}_i)
 [a_{y} ({\bf R}_i+{\bf z}) - a_{y} ({\bf R}_i-{\bf z})]
\nonumber\\
& & \qquad + \; \hbox{\rm cyclic permutations} \;.
\end{eqnarray}
However, in the AFD state, this term is zero.  To account correctly
for the coupling between ${\bf a}_i$ and ${\bf f}_i$, it is
necessary to include higher-order terms.  The lowest non-zero term
in the AFD state is quadratic in both ${\bf a}_i$ and ${\bf f}_i$.
Defining $w_{i,x}$ by
\begin{equation}
w_{i,x} = \frac{1}{8} \sum_{{\bf d}= \pm {\bf y},\pm {\bf z}}
  [a_x ({\bf R}_i+{\bf d}) - a_x ({\bf R}_i)]
\end{equation}
and similarly for $w_{i,y}$ and $w_{i,z}$, we can write
\begin{eqnarray}
 E_2^{\rm F-A} & = & \sum_i [ \; G_{xxxx} f_{i,x}^2 w_{i,x}^2 +
        G_{xxyy} f_{i,x}^2 (w_{i,y}^2 + w_{i,z}^2) \nonumber\\
& & \qquad + \; \hbox{\rm cyclic permutations} \;] \;.
\end{eqnarray}
(In principle a term $G_{xyxy} f_{i,x} f_{i,y} w_{i,x} w_{i,y}$
could also be included, but for practical reasons we have not done
so in this work.) In summary, up to the fourth-order terms, the
coupling between ${\bf a}_i$ and ${\bf f}_i$ is expressed as
\begin{equation}
E^{\rm F-A} = E_1^{\rm F-A} + E_2^{\rm F-A} \; .
\end{equation}

\section{First-principles calculations}

The expansion parameters in the model Hamiltonian can be obtained
from a set of first-principles calculations.  We use
density-functional theory within the local density approximation
(LDA). The technical details and convergence tests of the
calculations can be found in Ref.  \onlinecite{king1}. The use of
Vanderbilt ultra-soft pseudopotentials\cite{vand1} allows a
low-energy plane-wave cutoff to be used for first-row elements, and
also allows inclusion of semicore shells of the metal atoms.  This
makes high-accuracy large-scale calculations of materials involving
oxygen and $3d$ transition-metal atoms affordable.  A generalized
Kohn-Sham functional is directly minimized using a preconditioned
conjugate-gradient method.\cite{king1,payne,aria}  We use a (6,6,6)
Monkhorst-Pack k-point mesh\cite{monk} for single-cell
calculations (216 k-points in the full Brillouin zone), and the
corresponding smaller sets of mapped k-points for supercell
calculations.

\begin{table}
\caption{FE soft-mode eigenvectors for $AB$O$_3$ cubic perovskites
SrTiO$_3$, CaTiO$_3$, and NaNbO$_3$. ${\rm O}\parallel$ and
${\rm O}\perp$ indicate oxygen displacement parallel and
perpendicular to O---$B$ bond, respectively.
\label{t_fe}}
\begin{tabular}{lrrr}
                          &  SrTiO$_3$  &  CaTiO$_3$  &  NaNbO$_3$ \\
   $\xi_A$                &  0.472      &   0.698     &  0.449 \\
   $\xi_B$                &  0.612      &   0.330     &  0.625 \\
 $\xi_{{\rm O}\parallel}$ & -0.287      &  -0.157     &  -0.232 \\
   $\xi_{{\rm O}\perp}$   & -0.400      &  -0.436     &  -0.421
\end{tabular}
\end{table}

The calculation of expansion parameters related to the FE modes
follows the procedure presented in Ref.\onlinecite{zhong2b},
Sec.\ III. The soft mode eigenvectors for SrTiO$_3$, CaTiO$_3$, and
NaNbO$_3$ as calculated by King-Smith {\it et al.}, are summarized in
Table \ref{t_fe}. The calculated expansion parameters for the FE
modes are given in the top portion of Table \ref{t_para}.

\begin{table}
\caption{Expansion parameters of the Hamiltonian for SrTiO$_3$,
CaTiO$_3$, and NaNbO$_3$. Energies are in Hartree. FE local-mode
amplitudes are in units of lattice constant ($a$= 7.30a.u.,
7.192a.u., and 7.396a.u. for SrTiO$_3$, CaTiO$_3$, and NaNbO$_3$,
respectively); AFD local-mode amplitudes are in radians.
\label{t_para}}
\begin{tabular}{lcrrr}
           &            &  SrTiO$_3$  &  CaTiO$_3$  &  NaNbO$_3$ \\
\hline
 FE on-site   & $\kappa_2$ &  0.0559     &  0.0240     &  0.0679 \\
           & $\alpha$   &  0.150      &  0.023      &  0.168  \\
           & $\gamma$   &  -0.191     &  -0.006     & -0.256 \\
 FE intersite & $j_1$      &  -0.02034   &  -0.01186   & -0.02378\\
           & $j_2$      &   0.04274   &   0.02750   &  0.03078 \\
           & $j_3$      &   0.005722  &   0.002040  &  0.006460 \\
           & $j_4$      &  -0.003632  &  -0.002886  & -0.005446 \\
           & $j_5$      &   0.004882  &   0.001132  &  0.004820 \\
           & $j_6$      &   0.001416  &   0.000672  &  0.002358 \\
           & $j_7$      &   0.000708  &   0.000336  &  0.001178 \\
 FE dipole    & $Z^*$      &   8.783     &   6.768     &   9.179 \\
        & $\epsilon_{\infty}$ & 5.24  &   5.81      &   4.96  \\ \hline
 ADF harmonic  &$\kappa^A_2$&   0.162238  &   0.022244  &  0.095852 \\
           & $j_1^A$    &   0.010526  &   0.086972  &  0.034884 \\
           & $j_2^A$    &   0.000820  &   0.000544  &  0.002360 \\
           & $j_3^A$    &  -0.002782  &  -0.001398 &  -0.000272 \\
           & $j_4^A$    &  -0.105414  &  -0.112050 &  -0.097678 \\
           & $j_5^A$    &   0.009460  &   0.010792  &  0.009752 \\
           & $j_6^A$    &   0.002577  &   0.001262  & -0.000318 \\
           & $j_7^A$    &  -0.001288  &  -0.000632 &   0.000160 \\
           & $j_8^A$    &   0.013769  &   0.013956  &  0.014868 \\
 AFD 3rd Order & $B_3$      &   $-$       &   0.0056    &   0.0029  \\
 AFD 4th Order & $\alpha^A$ &   0.05433   &   0.04970  &   0.03775  \\
           & $\gamma^A$ &   0.04706   &   0.02414  &   0.04301  \\
\hline
Elastic    & B$_{11}$   &   5.14      &   5.15     &    6.63  \\
           & B$_{12}$   &   1.38      &   1.22     &    0.96  \\
           & B$_{44}$   &   1.56      &   1.29     &    1.07  \\
\hline
 FE--E 
coupling   & B$_{1xx}$  &   -1.41     &   -0.59    &    -1.71 \\
           & B$_{1yy}$  &    0.06     &   0.06     &    0.50  \\
           & B$_{4yz}$  &   -0.11     &   -0.10    &    0.00  \\
 AFD--E
coupling   &B$^A_{1yyx}$&   0.260     &   0.234    &    0.316 \\
           &B$^A_{2yyx}$&   -0.068    &   -0.008   &    0.026 \\
           &B$^A_{3yyx}$&   0.000     &   -0.034   &    0.031 \\
           &B$^A_{4yzx}$&   0.044     &   0.040    &    -0.041 \\
 FE--AFD
coupling   & G$_{xy}$   &   0.0061    &   -0.0001  &    0.0014 \\
           & G$_{xxxx}$ &     0.53    &   0.72     &   0.35  \\
           & G$_{xxyy}$ &     0.11    &   0.29     &   0.06
\end{tabular}
\end{table}

The calculation of the AFD expansion parameters follows a similar
procedure as for the FE ones.  The AFD eigenvector itself does not
need to be calculated, since it is determined by symmetry.  The LDA
total energy vs. AFD distortion, with polarization along $x$ and
$z$, and at k-points X=($\pi/a$,0,0), M=($\pi/a$,$\pi/a$,0), and
R=($\pi/a$,$\pi/a$,$\pi/a$), are calculated.  The arrangements of
the AFD local modes are the same as for the FE-mode calculations
as shown in Figs.~3(a)--(f) of Ref. \onlinecite{zhong2b}.
However, the arrangements at the $\Gamma$ point (Fig.~3(a) of Ref.
\onlinecite{zhong2b}) and at the X point (Fig.~3(b) of Ref.
\onlinecite{zhong2b}) involve no actual distortions.  So the work
reduces to
four 10-atom cell calculations and two 20-atom cell calculations
needed to determine $j_1^A$, $j_2^A$, $j_3^A$, $j_4^A$, $j_6^A$,
and $j_5^A+2j_7^A$.  The decomposition of $j_5^A$ and $j_7^A$
follows the same argument as for the FE case.  It is difficult to
perform sufficient LDA calculations to carry out the decomposition,
and probably not very important to do so.  Instead, we rely on the
heuristic that the interaction between two AFD local modes should
be minimal (in practice, zero) when they are so arranged that
reversing one of them induces a minimum bond-length change.  For
the AFD case, this leads to $j_6^A+2j_7^A =0$, allowing $j_5^A$ and
$j_7^A$ to be obtained separately. The fourth-neighbor interaction
parameter $j_8^A$ is obtained from LDA calculations on a 15-atom
supercell.

As mentioned before, we need to include the effect of third-order
intersite coupling in the effective Hamiltonian in some compounds
having large AFD distortions.  This kind of interaction generates a
coupling among three distortions: an R-point mode with polarization
(110), an M-point ($\pi/a$,$\pi/a$,0) mode polarized along (001),
and an X-point (0,0,$\pi/a$) mode polarized along (110). To
determine the strength of this coupling, we carry out a calculation
with the above R-point and M-point distortions frozen in (20-atom
supercell), and calculate the forces of X-point character.  We
find that the force on the A-metal atom (Ca or Na) is non-zero, and
opposite in sign to the force on the oxygen atom.  This is in
qualitative agreement with the experimentally observed
displacements in the low-temperature phase, which are also opposite
in sign.  We then calculate the projection of these forces onto the
X-point mode, under the simplifying assumption that the latter
consists of equal-amplitude out-of-phase motion of the two atoms.
This projection determines our third-order intersite coupling
parameter $B_3$.

The calculation of the anharmonic coefficient $\alpha^A$ is
performed with an R-point distortion polarized along (001).
Typically a set of eight calculations are performed for each
compound.  The resulting LDA total energy is fitted to a polynomial
$E_0 + c_2 a^2 + c_4 a^4$ using standard least-squares methods to
extract $\kappa_2^A$ and $\alpha^A$.  A single R-point calculation
with polarization in the (111) direction is used to extract
$\gamma^A$.

The four parameters describing the coupling between AFD modes
and elastic strains are obtained by performing
four more 10-atom supercell calculations:
at the M point, ${\bf a}\parallel(001)$,
   with an isotopic strain $\eta_1=\eta_2=\eta_3=0.01$;
at the M point, ${\bf a}\parallel(001)$,
   with strain $\eta_1=0.01$;
at the X point, ${\bf a}\parallel(001)$,
   with strain $\eta_1=0.01$; and
at the R point, ${\bf a}\parallel(111)$,
   with strain $\eta_4=\eta_5=\eta_6=0.01$.
Extra care has been taken to ensure cancellation of errors due to
k-point sampling and basis-size differences for the different unit
cells involved.

Finally, the couplings between FE and AFD modes are determined as
follows.  The harmonic coupling $G_{xy}$ is determined by
considering a geometry in which the primitive cell has been
tripled along the $x$ direction, and for which $f_x$ is non-zero in
two primitive cells and $a_y$ is non-zero in the third.  The
anharmonic couplings $G_{xxxx}$ and $G_{xxyy}$ are determined from
a series of calculations on a doubled-cell configuration in which
${\bf f}(\Gamma)$ $\parallel$ (100) and ${\bf a}(R)$ $\parallel$
(100) or (010).

The calculated parameters for all three compounds are listed in
Table \ref{t_para}.  We note that the intersite interaction
parameters between AFD local modes have a much stronger anisotropy
than those between FE modes.  For FE, the $j$'s show no marked
anisotropy.  (Of course, when the Coulomb interaction is included,
the actual interaction between FE modes are quite anisotropic). On
the other hand, for the first-neighbor AFD couplings, $j_1^A$ is
more than one order of magnitude larger than $j_2^A$, which is
reasonable since $j_2^A$ involves no distortion of oxygen
octahedra.  For second neighbors, $j_4^A$ is again much larger than
the others, confirming that the distortion of oxygen octahedra
dominates the energy for AFD distortions.  This observation that
the in-plane interaction parameters are much stronger than the
out-of-plane ones is what prompted us to include also the
fourth-neighbor in-plane AFD interactions in the effective
Hamiltonian.

\section{Results}

After the expansion parameters have been determined from
first-principles calculations, the finite-temperature properties of
the compounds can be calculated using Metropolis Monte Carlo (MC)
simulations.\cite{MC}  The details of the MC simulations involving
FE and elastic distortions have been described in Sec. IV of Ref.
\onlinecite{zhong2b}.  With the AFD distortions included, the
number of degrees of freedom is increased from 6 to 9 per unit
cell.  All the details are very similar, the main difference being
that the AFD degrees of freedom introduce many more possibilities
for modes which may go soft.  The primary candidates for soft AFD
modes are three modes at the R point and one mode at each of the
three M points, all of them involving only rigid rotations of
oxygen octahedra.  We thus have to consider a rather complicated
set of order parameters, and we anticipate that complex phases may
form.

The results for the three different compounds SrTiO$_3$, CaTiO$_3$,
and NaNbO$_3$ will be presented in the three following subsections.
Because of the strong sensitivity of structural phase transition
temperatures to the lattice constant and the well-known $\sim1\%$
LDA underestimate of lattice constants, we concentrate on presenting
calculated transition temperatures and transition sequences at
the experimental lattice constants.  We thus implicitly apply a
negative fictitious pressure to the simulation cell, as explained
in Sec.\ IV of Ref. \onlinecite{zhong2b}.

\subsection{SrTiO$_3$}

\begin{figure}
\epsfysize=4.5in
\centerline{\epsfbox{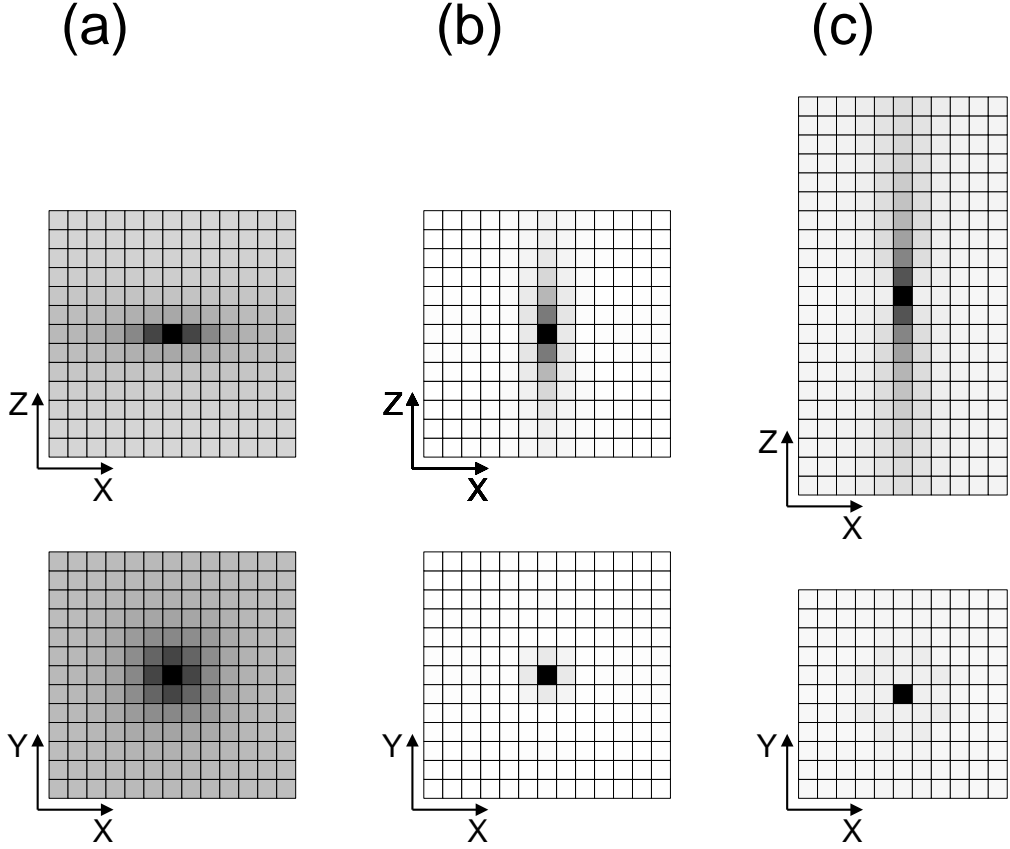}}
\medskip
\caption{
(a) Intersite correlations of AFD local modes $\langle
a_z(0,0,0) a_z(x,y,z)\rangle $ in SrTiO$_3$ at $T$=150K.  The
correlations in the $x$-$z$ plane ($y=0$, top) and the $x$-$y$ plane
($z=0$, bottom) are shown.  Each small square represents one lattice site;
the origin lies at the center.  Grey scale denotes the degree
of correlation: black for perfect correlation, white for
zero correlation.
(b) Same for FE local-mode correlations
$\langle f_z(0,0,0) f_z(x,y,z)\rangle$.
(c) Corresponding FE local-mode correlations
in BaTiO$_3$ at $T$=320K.
%
\label{corr}}
\end{figure}

Thermodynamic properties for this compound have been calculated
and published in Ref. \onlinecite{zhong3}.  A pressure $P_0 =
-5.4$GPa is applied to restore the experimental lattice constant.
A transition from the cubic phase to a tetragonal AFD structure
at 130K, and two further FE phase transitions at 70K and 10K,
were predicted.  A later quantum path-integral MC simulation
revealed that quantum fluctuations suppress the FE phases entirely,
and reduce the AFD phase transition temperature to
110K.\cite{zhong4} This gives excellent agreement with experiment,
which reveals a single AFD phase transition at 105K,\cite{Tc} and no
unambiguous phase transition (but the presence of ``quantum
paraelectric'' behavior \cite{muller,viana,mart}) at low
temperature.  Our calculated pressure-temperature phase diagram
showed that the FE and AFD instabilities have opposite trends with
pressure, and FE and AFD instabilities tend to suppress each
other.

We have performed some further simulations to investigate the
behavior of the AFD and FE local modes, and in particular the
nature of the intersite correlations for FE and AFD local modes
in the cubic phase but just above the phase transition temperature.
The M-point AFD modes do not appear to be important for SrTiO$_3$,
so we focus on the two vector order parameters ${\bf f}(\Gamma)$
and ${\bf a}({\rm R})$ associated with the zone-center FE modes
and the zone-corner AFD modes.  Since the order parameters are
vectors, the correlation functions are second-rank tensors.
These can be calculated from our simulations as
\begin{equation}
S_{\alpha\beta}(x,y,z) = \langle v_\alpha (x_0,y_0,z_0)
   v_\beta (x_0+x,y_0+y,z_0+z) \rangle \; ,
\end{equation}
where the average is taken over all the sites $x_0, y_0, z_0$ in
the MC simulation cell and over all MC sweeps $t$. Here the
$v_\alpha$ denote the components of the FE or AFD order parameters
[$f_\alpha$ for FE, $a_\alpha(-)^{l+m+n}$ for AFD, where ${\bf
R}_i=l{\bf x}+m{\bf y}+n{\bf z}$].  We can get a good picture of
the nature of the correlation by investigating the diagonal
elements ($\alpha=\beta$) only.  Since the three Cartesian
directions are equivalent, it suffices to present $S_{zz}(x,y,z)$.

In Fig.~\ref{corr}(a), we show the calculated AFD correlation
function $S^{\rm AFD}_{zz}(x,y,z)$ in the two planes $x$-$z$ and
$x$-$y$ for SrTiO$_3$ at T=150K, in the cubic phase but just
above the AFD phase transition temperature of 130K.  We find
that the correlations are quite strong in $x$-$y$ plane, with a
correlation length of about three lattice constants.  Along the
$z$ direction, even the nearest-neighbor vectors are almost
completely uncorrelated.  Thus, the shape of the ``equal correlation
surface'' for AFD local modes is disc-like.  This is easy to
understand on the basis of the RUM picture.\cite{dove} Since the
AFD local modes involve a rotation of oxygen octahedra, and any
distortion of the oxygen octahedra involves a large energy cost
(as shown by the large magnitude of $j_1$, $j_4$, and $j_8$),
the AFD octahedral rotations about $\hat{\bf z}$ correlate strongly
in the $x$-$y$ direction.  On the other hand, the rigidness of
the octahedra does not impose any relation between $z$-polarized
AFD modes in different $z$ planes (as reflected in the very small
$j_2$).  Thus, the pancake-like correlation naturally results.

Fig.~\ref{corr}(b) shows the corresponding FE correlation
function $S^{\rm FE}_{zz}(x,y,z)$ for SrTiO$_3$ at T=150K (the
FE phase transition occurs at 70K).  Its behavior is just the
reverse of the AFD modes, being strong along the $z$ direction
and weak in the $x$-$y$ plane, and resulting in a needle-shaped
``equal correlation surface.'' This behavior is a direct consequence
of the anomalously large mode effective charges in the cubic
perovskite compounds\cite{zhong1} which strongly suppress the
longitudinal FE fluctuations and leads to the strong correlation
of $f_z$ in the $z$ direction.  On the other hand, the transverse
FE modes can easily go soft, resulting in a short correlation
length for $f_z$ in the $x$-$y$ plane.

The above picture of the correlation functions for AFD and FE
local modes are presumably quite general for the cubic perovskite
compounds.  For the FE modes, we decided to repeat the calculations
for the case of BaTiO$_3$, where the AFD instability does not
intervene.  We show in Fig.~\ref{corr}(c) the correlation function
calculated in the cubic phase at T=320K, about 20K above the FE
phase transition temperature.  We can see the behavior of the
correlations is again needle-like and even more pronounced than
for SrTiO$_3$, presumably because we are closer to the FE transition
temperature.  (An elongated supercell was used to accomodate the
correlations in this case.)

\begin{figure}
\epsfysize=3.0in
\centerline{\epsfbox{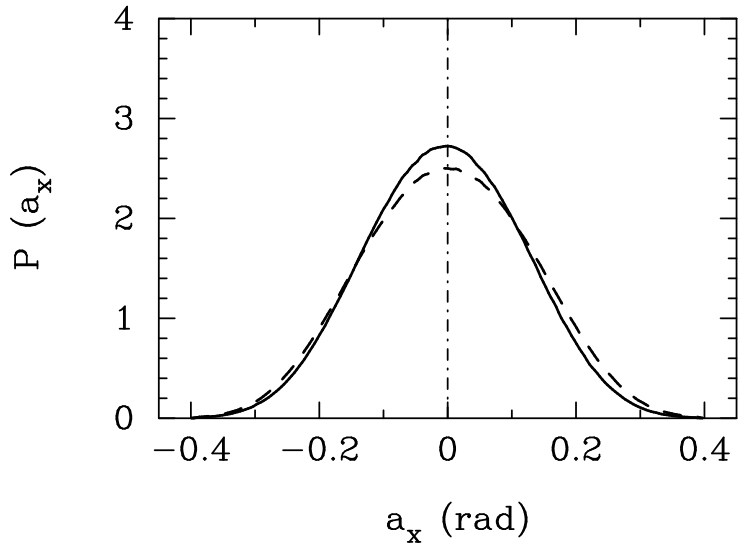}}
\medskip
\caption{The probability distribution of the Cartesian component of
the local-mode variable $a_x$ for SrTiO$_3$ in the cubic phase at
$T$=140K (dashed line) and $T$=150K (solid line).
\label{sr_dos}}
\end{figure}

As was done for the FE modes in BaTiO$_3$,\cite{zhong2} it is
revealing to compute the equilibrium distribution of one
cartesian component of the AFD order parameter ${\bf a}({\rm R})$
in the cubic phase just above the phase transition temperature
in SrTiO$_3$.  This is shown in Fig.~\ref{sr_dos}, where it
can be seen that the distribution looks approximately
Maxwellian.  This is indicative of a transition having a character
much closer to the displacive than to the order-disorder limit.

\subsection{CaTiO$_3$}

CaTiO$_3$ is one of the more complicated perovskites.
Experimentally, it is found to have two stable phases,
an orthorhombic phase at lower $T$ and a cubic phase above
$1530$K.\cite{granicher}  Some recent experiments suggest that
the transition is to a highly disordered cubic phase.\cite{vogt}
The room-temperature orthorhombic phase has a very complicated
structure with a 20-atom unit cell.  The displacements of all
the atoms away from their idea positions have been determined in
Ref.\ \onlinecite{kay}.  The refined structure as a function of
temperature has also been determined recently using X-ray
diffraction.\cite{liu,sasaki} This complicated structure can be
decomposed into a simultaneous freezing in of three AFD modes:
an R-point mode polarized along (1$\bar1$0) with rotation angle
0.20 (angles in radians), an M-point mode
polarized along (001) with rotation angle 0.14, and an X-point
mode polarized along (1$\bar1$0).  The X-point mode involves not
only the rotation of oxygen octahedra, but also an associated
displacement of Ca atoms.  The ratio of O and Ca displacement is
about $1:-3$ and the oxygen octahedral rotation angle is only
about 0.03.

For such a complicated structure, even a complete first-principles
determination of its $T=0$ structure would be very difficult.
However, we can arrive at a partial understanding of this structure
as follows.  Our calculations show that the reference cubic
structure is unstable towards either the R-point or the M-point
AFD mode individually (negative $\omega^2$), whereas it is {\it
stable} with respect to the X-point mode individually (positive
$\omega^2$, no double-well behavior).  In fact, the X-point mode
involves a strong distortion of the oxygen octahedron, and thus
is far from soft.
However, the symmetry of the crystal is such that if both the
R-point and M-point AFD mode distortions are already simultaneously
present, then the Ca and O atoms experience forces in the pattern
of the X-point mode, as a result of the cubic anharmonic
interactions discussed in Sec.\ IIB.  Thus, under these conditions
the crystal would necessarily acquire some X-point mode distortion.
We therefor conclude that the appearance of the X-point mode
distortion must be the result of the third-order coupling between
R-, M-, and X-point AFD modes.  This is a major reason why we
chose to include the third-order coupling in our model Hamiltonian.
The strain coupling is expected to be important in the determination
of the actual magnitude of the X-mode distortion, complicating the
problem and making a complete quantitative LDA determination of
the X-mode amplitude difficult.

\begin{figure}
\epsfysize=3.0in
\centerline{\epsfbox{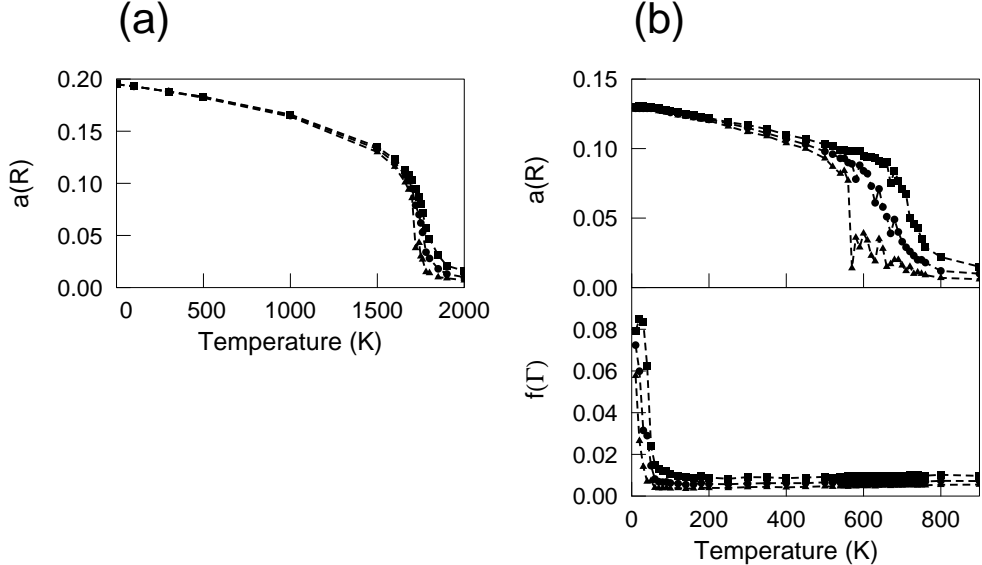}}
\medskip
\caption{
(a) Calculated AFD order parameters $a(R)$ of CaTiO$_3$
vs.\ $T$ at $P_0 = -11.3$GPa.
(b) Calculated FE and AFD order parameters of NaNbO$_3$
vs. $T$ at $P_0 = -4.3$GPa.
\label{order}}
\end{figure}

To obtain the phase transition sequence, we start the MC simulation
at a high temperature ($T>2000$K) and equilibrate the system for
10,000 MC sweeps.
An isotropic pressure $P_0=-11.3$GPa is imposed to restore the
experimental lattice constant, and all the subsequent simulations
are done under this pressure.  The temperature is reduced in
small steps (as small as 10K around the transition temperature)
with 30,000 MC sweeps at each $T$ to ensure equilibration.  The
order parameters are accumulated over the last 20,000 MC sweeps,
after checking that they do not vary significantly over this
period.  In our simulation, we find that except for the R-point
AFD order parameters ${\bf a}({\rm R})$, all other FE or AFD
order parameters are zero throughout the simulation.  So the only
phase transition we observe is associated with ${\bf a}({\rm R})$.
Fig.~\ref{corr}(a) shows the calculated order parameters ${\bf a}
(R)$ as a function of temperature.  (What we actually plot are
the averaged maximum, intermediate, and minimum absolute values
of the order-parameter components.) At high temperature ($T>1800$K)
the system is in the cubic structure, where all the components
are zero.  The material goes through a phase transition at 1750K,
where all three order parameter components increase simultaneously.
Our phase transition temperature is close to the experimentally
measured 1530K, but we obtain the wrong low-temperature structure.
Ours is rhombohedral with a 10-atom cell, instead of orthorhombic
with a 20-atom cell as observed experimentally.

The difference between our theoretical result and experiment is
not quite as dramatic as it might seem.  Both the structure and
the structural energy are very similar for the R-point and the
M-point AFD mode polarized along (001).  Note that in the observed
structure, the amplitude of the M-point mode is about $\sqrt{2}/2$
times the the amplitude of the R-point mode polarized in the
(1$\bar1$0) direction.  Thus, we can say that the main difference
between our calculated structure and the observed one is that
one component of the R-point AFD mode is replaced by an M-point
mode in the observed structure.  As argued above, the additional
presence of the X mode is just a result of third-order anharmonic
coupling.  In fact, we find that if we artificially increase the
third-order coupling constant $B_3$ by a factor of 5, we recover
the experimental $T=0$ structure in our MC simulations.

Clearly, the fact that the structure is strongly affected by
relatively weak anharmonic intersite interactions makes the
determination of the correct low-temperature phase very difficult
in CaTiO$_3$.  It is possible that a more careful treatment of
the cubic intersite interactions (for example, an independent
determination of the coupling constants associated with all three
of the cubic anharmonic invariants discussed in Sec.~II.B, or
three-site or further-neighbor terms) might bring a better
agreement with experiment, although one should not rule out
the possibility that neglect of quantum fluctuations\cite{zhong4}
or intrinsic limitations of the LDA might be at fault.

\subsection{NaNbO$_3$}

Experimentally, NaNbO$_3$ is probably the most complex cubic
perovskite known.  The high-temperature phase is the simple
prototype cubic structure as in the other cubic perovskites.
Below 910K, a whole series of structural phase transitions has
been found and at least six more phases have been identified. As
the temperature decreases, the compound first goes through a
cubic--tetragonal transition at 910K with freezing in of ${\bf
a}({\rm R})$ modes polarized along one axis.  There are then
three orthorhombic phases present in the temperature range
845--638K, the most complicated having a unit cell containing
24 NaNbO$_3$ formula units.  All of these phases can be
regarded as given by rigid rotations of oxygen octahedra,
accompanied by small induced X-point distortions.  From 638K
down to at least 170K, NaNbO$_3$ is antiferroelectric with an
orthorhombic unit cell containing eight formula units.
At even lower temperature, the crystal has been reported to
transform into either a rhombohedral \cite{landolt}
or monoclinic\cite{lines} structure.

The complexity of the structural phase-transition sequence suggests
the presence of several competing structural instabilities with
very similar free energies.  In principle, all the distortions
involved in the observed structures of NaNbO$_3$ are included in
our model. However, it is not realistic to expect that the
calculated structural energies and free energies will be in
exactly the right order, given the complexity of the problem and
the level of accuracy of current first-principles based approaches.
Nevertheless, we believe a first-principles study of NaNbO$_3$
is still important in identifying the most prominent distortions,
as well as for demonstrating the limitations of such approaches.

The determination of the structure is done using MC simulations
on a cubic $ 12 \times 12 \times 12 $ simulation supercell.  An
isotropic pressure $P_0=-4.3$GPa is imposed to restore the
experimental lattice
constant, and all the subsequent simulations are done under this
pressure.  We start the simulation at very high temperature and
equilibrate.  The temperature is reduced in small steps ranging
from 10K to 50K depending on proximity to a phase transition.
At each temperature step, at least 40,000 MC sweeps are used to
ensure that equilibrium is reached.  The order parameters are
accumulated over the last 30,000 MC sweeps.

The calculated averages of order parameters ${\bf a} ({\rm R})$ and
${\bf f}(\Gamma)$ are shown in Fig.~\ref{corr}(b) as a function
of temperature.  All other modes are found to be zero throughout
the simulation.  As was the case for CaTiO$_3$, the averaged
maximum, intermediate, and minimum absolute-value components are
plotted.  At high temperature ($T>800$K), the system is in the
cubic structure with all the order-parameter components close to
zero. As $T$ decreases to about 700K, one AFD component increases
rapidly and becomes significantly non-zero, and the structure
transforms from cubic to tetragonal.  With further decrease of
temperature, a second component became non-zero, indicating the
occurrence of an orthorhombic phase.  Below 560K, a third AFD
component grows and the structure becomes rhombohedral.  At very
low temperature (below 50K), the simulation also apparently shows
a sequence of three ferroelectric transitions, and the compound
ends up in a rhombohedral ferroelectric structure at very low
temperature.

Our first cubic--tetragonal phase transition compares favorably
with experiment; we obtain the correct structure and underestimate
the transition temperature by only $\sim$20\%.  In the orthorhombic
phase, however, the calculated structure is much simpler than
the observed one.  Only one orthorhombic phase seems to occur in
our simulation.  However, Fig.~\ref{corr}(b) shows signs of
fluctuations occurring in the vicinity of this phase (these
fluctuations persist even if the number of MC sweeps is increased
significantly).  This indicates that the orthorhombic phase is
not very stable, and may involve a mixing of different phases.
Moreover, the transitions do not appear very distinct in
Fig.~\ref{corr}(b) as a result of finite-size broadening, so
increasing the lattice size may help to resolve the different
phases.  However, the computational load increases rapidly with
increasing lattice size, and it becomes impractical to carry out
simulations at much larger size.  Our inability to get the correct
AFE phase at room temperature is probably the most significant
failure of our approach.  Our zero-temperature structure is
ferroelectric, but since the FE phases occur only at such low
temperatures, it is likely that quantum fluctuations would need
to be included to determine the actual low-temperature
structure.\cite{zhong4}

\section{Discussion and Conclusion}

In this and previous studies, we perform a series of {\it ab
initio} studies of the thermodynamic properties of perovskite
compounds.  Without introducing any adjustable parameters, we
have calculated structural transition sequences, transition
temperatures, phase diagrams, and other thermodynamic properties
based on first-principles calculations.  For compounds with simple
phase transitions, like BaTiO$_3$ and SrTiO$_3$, our calculated
thermodynamic properties agree very well with experiment
observations. For more complicated compounds like CaTiO$_3$ and
NaNbO$_3$, our results are less satisfying.

There are two major sources of errors, the inaccuracy of LDA
calculations and the imperfection of our models.  Our LDA
calculations have been carefully performed to avoid possible
errors, and convergence has been carefully tested.  As for the
intrinsic accuracy of LDA, our calculated structural parameters
and energies are within a few percent of experimental values.
Although this is the usual high accuracy observed generally for
the LDA, it is unfortunately not enough for truly accurate
determination of the thermodynamic properties of perovskites.
For example, it is embarrassing that we are forced to choose
between carrying out the calculations at the theoretical equilibrium
lattice constant or the experimental lattice constant (negative
fictitious pressure); this choice can affect phase transition
temperature by $\sim$100\%.  We regard this as being the most
important probable source of error in our calculations.

It is also possible to improve our model Hamiltonian.  For example,
our restricted assumption for the form of the third-order intersite
interactions may be lifted, resulting in a significantly more
complicated model Hamiltonian. Also, other higher-order terms
can be included in the Hamiltonian.  It would also be possible
to include more degrees of freedom per cell, or include eigenvector
information from more k-points of the Brillouin zone when defining
the local-mode vectors, to treat other phonon excitations more
accurately.\cite{rabe-w}  However, in view of the current accuracy
of first-principles calculations, we are not sure that these
modifications would dramatically improve our results.

Finally, we emphasize that all of the MC simulations reported here
treat the atomic motion purely classically.  As mentioned above, we
have recently reported results of quantum path-integral MC
simulations showing that quantum fluctuations of the atomic
coordinates (i.e., zero-point motion) can shift transition
temperatures by tens of degrees, and in some cases even eliminate
delicate phases.\cite{zhong4}  Certainly this remains an important
avenue of investigation for CaTiO$_3$ and NaNbO$_3$, but we
nevertheless think it unlikely that inclusion of quantum
fluctuations would immediately resolve the discrepancies with the
experimental phase diagrams for these compounds.

In conclusion, we have extended our previous first-principles
theory for perovskite ferroelectric phase transitions to treat
also antiferrodistortive transitions. We apply this approach to
the three cubic perovskite compounds SrTiO$_3$, CaTiO$_3$, and
NaNbO$_3$, and calculate their thermodynamic properties including
phase transition sequences and transition temperatures.  For
SrTiO$_3$, our calculated results are in good agreement with
experiment.  For CaTiO$_3$ and NaNbO$_3$, our calculated structural
transitions have the correct general trend and the transition
temperatures are in rough agreement with experiment, but the
calculated transition sequences are not correct in detail.  We
attribute this to the larger distortions and many multiple
competing instabilities in these compounds.  For SrTiO$_3$, we
also analyzed the intersite correlations for both FE and AFD
local modes, finding needle-like and pancake-like correlations
respectively for FE and AFD modes as expected on physical grounds.

\acknowledgments

This work was supported by the Office of Naval Research under
contract number N00014-91-J-1184.   Partial supercomputing support
was provided by NCSA grant DMR920003N.

\end{document}